\let\texyear\year
\documentclass{IEEEtran}
\usepackage{cite}
\usepackage{amsmath,amssymb,amsfonts}
\usepackage{algorithmic}
\usepackage{graphicx}
\usepackage{textcomp}
\usepackage{multirow}
\usepackage{url}
\usepackage{hyperref}
\let\ieeeaccessyear\year
\let\year\texyear
    \usepackage{orcidlink}
\let\year\ieeeaccessyear
\definecolor{accessblue}{RGB}{0,105,154}

\begin{document}
\title{FPGA-Accelerated Neuromorphic Vision System for Real-Time Orbital Object Detection}
\author{
Diego Hernández\,\orcidlink{0009-0002-6540-2257},
Sebastián Valdivia\,\orcidlink{0009-0005-0657-2303},
Vicente Westerhout\,\orcidlink{0009-0000-6363-7859},
Esteban~Vera\,\orcidlink{0000-0001-8387-8131} and
Daniel Yunge\,\orcidlink{0000-0001-7149-2768}
\newline Pontificia Universidad Católica de Valparaíso, Valparaíso, Chile
\newline e-mail: diego.hernandez.v@pucv.cl; sebastian.valdivia@pucv.cl; vicente.westerhout@pucv.cl; esteban.vera@pucv.cl; daniel.yunge@pucv.cl
}




\maketitle
\begin{abstract}
The escalating congestion in orbital space demands advanced monitoring solutions. This work presents a comprehensive open-source framework for neuromorphic resident space object (RSO) detection, adapting the foundational grid clustering algorithm for FPGA acceleration. The system integrates a single event-based camera (EBC) with a custom, distributed processing architecture, where rapid spatial quantization is executed in programmable logic (FPGA) and cluster formation is managed by a software client. We validate this architecture through systematic sampling of night-sky observations from the EVAS dataset, demonstrating 97\% detection accuracy for RSOs. The implementation, which serves as a foundational toolkit for event-based FPGA processing, achieves efficient throughput with a total power consumption of 8.5 W and deterministic processing latencies below 62 ms. The architecture's energy efficiency and high-precision detection position it as a viable solution for distributed space surveillance networks.
\end{abstract}

\section{Introduction}
\label{sec:introduction}
The rapidly escalating congestion in orbital space presents critical challenges for global space agencies, with current estimates indicating over 130 million resident space objects (RSOs) in Earth's orbit\cite{ESA2023}. High-velocity collision risks, exemplified by the 2009 Iridium–Cosmos incident \cite{NASA2009}, underscore the urgent need for advanced monitoring solutions capable of detecting small, fast-moving RSOs in real time. Traditional space surveillance approaches face fundamental limitations in this domain, particularly regarding power consumption, temporal resolution, and operational flexibility. Ground-based radar systems, while effective for long-range detection and capable of day-and-night operation, often require high transmitted power, high-gain antennas, and substantial infrastructure for RSO detection and tracking \cite{Choi2017RadarSSA,Bira2025CheiaRadar}. These requirements make them less suitable for compact, low-power distributed sensing nodes. Conventional frame-based CMOS cameras, in turn, are constrained by illumination conditions and by readout mechanisms that can introduce motion blur or rolling-shutter distortions when imaging fast-moving objects, while their dynamic range is typically lower than that of event-based sensors \cite{Wilburn2004DenseCameraArray,Gallego2020,Rebecq2021HighSpeedHDR}.

Recent advances in event-based cameras (EBCs) and neuromorphic vision technologies represent a paradigm shift from conventional frame-based imaging, offering significant advantages in temporal resolution, dynamic range, and power efficiency. These bio-inspired sensors operate asynchronously, reporting pixel-level brightness changes as discrete events rather than capturing full frames at fixed intervals \cite{Gallego2020}. This approach enables microsecond temporal resolution, dynamic range exceeding 120 dB, and inherent data sparsity—characteristics particularly well-suited for space surveillance applications where detecting faint, fast-moving RSOs against static star fields is essential \cite{Posch2014}. The event-driven nature of neuromorphic vision reduces data volume by orders of magnitude compared to conventional video streams, addressing fundamental bandwidth and storage constraints in distributed monitoring networks.
The computational requirements of processing event-based data streams have driven innovation in specialized hardware architectures. FPGA-based implementations offer the flexibility to create custom processing pipelines that align with the unique characteristics of neuromorphic vision, enabling more efficient implementation of bio-inspired algorithms \cite{Karamimanesh2025}. The synchronous, frame-oriented processing paradigm of conventional processors fundamentally mismatches the asynchronous, sparse nature of event-based data, leading to inefficiencies in both performance and energy consumption.
This work presents a specialized implementation of the foundational grid clustering algorithm \cite{Schikuta1996} optimized for RSO detection through FPGA acceleration. The system architecture strategically leverages the parallel processing capabilities of programmable logic to execute spatial quantization operations with minimal latency and power overhead. The implemented system was evaluated using a single EBC–FPGA configuration, providing a foundational unit for scalable deployments. Extensive validation using both standardized datasets and field observations demonstrates the system's capability to reliably track RSOs with high accuracy while operating within strict power constraints, addressing critical needs in next-generation space situational awareness (SSA) systems.
A key contribution of this work lies in its use of a commercially available, off-the-shelf FPGA board, which makes the solution accessible, scalable, and cost-effective. Unlike custom hardware solutions that require specialized manufacturing, our approach leverages standard components, allowing easy replication and deployment in arrays of EBCs. This modularity enables a straightforward expansion to multi-EBC networks, where each FPGA handles one EBC, preserving deterministic performance and predictable power budgets. The low power consumption (8.5 W total) and compact form factor further broaden the applicability beyond space surveillance to robotics, aerial vehicles, and autonomous driving systems, where low weight and energy efficiency are critical.
\section{Related Work}
\label{sec:related}
Event-based vision has been explored for SSA in recent years. Afshar et al.\cite{Afshar2020} demonstrated the use of EBCs for RSO detection and tracking, achieving high temporal resolution but relying on software-based processing that limited throughput. More recently, Valdivia et al. introduced ARACHNID\cite{ARACHNID}, a neuromorphic EBC array system that shifts this processing paradigm by combining multiple EBCs with spiking neural networks for real-time detection and tracking of RSOs. In parallel, the EVAS dataset\cite{Valdivia2023} provided several recordings for RSO tracking with EBCs, highlighting the need for processing pipelines with lower power consumption and more deterministic latency than conventional CPU- or GPU-based approaches. Building on this line of work, the system proposed in this paper was integrated with the ARACHNID EBC array to evaluate its operation under real event-based observations, as further detailed in Section~\ref{sec:results:field}.
FPGA acceleration for event-based processing has been investigated primarily for neural network inference \cite{Karamimanesh2025} and low-level feature extraction. However, dedicated hardware implementations of clustering algorithms for neuromorphic data remain scarce. The work of Shi et al. \cite{Shi2014} accelerated DBSCAN on FPGA but required $O(n\log n)$ complexity and significant memory resources, limiting its applicability to high-rate event streams. In contrast, grid-based clustering \cite{Schikuta1996} offers $O(n)$ complexity and minimal state, making it inherently suitable for streaming hardware acceleration.
Our work distinguishes itself by providing a complete, open-source framework that couples an EBC with a low-cost FPGA board for real-time RSO detection. The hybrid client-server architecture partitions the algorithm to maximize hardware efficiency while maintaining flexibility, and the comprehensive statistical validation using information-theoretic metrics establish an optimal operating point for detection accuracy.
\section{Methodology}
\label{sec:methodology}

\subsection{System Architecture}

\begin{figure}[t]
 \centering
 \includegraphics[width=0.5\textwidth]{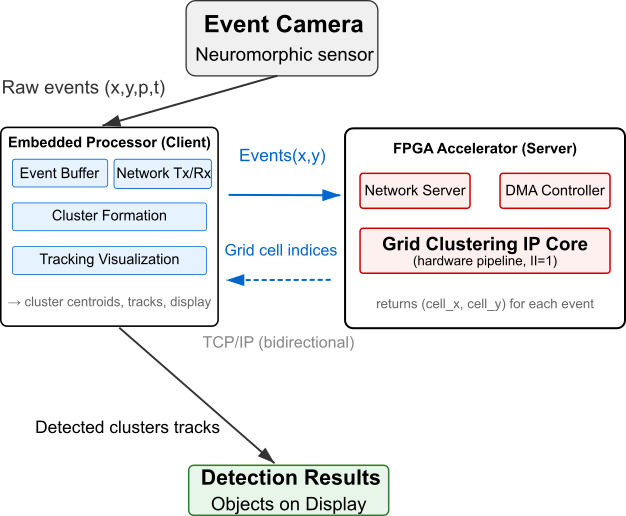}
 \caption{Diagram of the distributed processing architecture showing data flow between embedded processor (client) and FPGA accelerator (server).}
\label{fig1}
\end{figure}

The implemented system employs a client–server architecture designed around a single EBC and an FPGA processing unit, as illustrated in Fig.~\ref{fig1}. The client subsystem, implemented in Python and executing on an embedded processor, interfaces directly with an EBC through a USB 3.0 connection. This component handles real-time event capture and preprocessing, including spatial ROI filtering and removal of persistent events, with configurable parameters typically set to [20, 20, 580, 420] pixels for general sky coverage. The client maintains a dynamic event buffer that aggregates incoming events until either temporal (20,000 $\mu$s) or size (250 events) thresholds are met, implementing efficient data reduction strategies that preserve essential spatial and temporal information while minimizing computational overhead.
The hardware accelerator is built around a low-cost, off-the-shelf FPGA board that executes the computationally intensive grid quantization operations in programmable logic. The design integrates a direct memory access (DMA) controller for efficient data transfer and a custom grid clustering IP core. DMA channels transfer event coordinates between the processing system and programmable logic, minimizing software overhead and enabling deterministic processing times essential for real-time operation. The grid size is fixed to $16 \times 16$ to balance the trade-off between spatial resolution and computational overhead.
\subsection{Algorithm selection}
The selection of a clustering algorithm for real-time neuromorphic event streams is constrained by the need for low latency, deterministic execution, and minimal memory footprint. Table~\ref{tab:algo_comparison} compares three common clustering approaches. K-Means requires prior knowledge of the number of clusters $k$ and iterative refinement, making it unsuitable for streaming data with unknown object counts. DBSCAN, while robust to noise, demands $O(n\log n)$ or $O(n^2)$ neighborhood queries and substantial memory to maintain density connectivity. Grid-based clustering \cite{Schikuta1996} operates in a single pass over the data with $O(n)$ complexity, mapping each event directly to a discrete grid cell via simple arithmetic division. This stateless, parallelizable operation aligns perfectly with FPGA streaming architectures, enabling a pipelined implementation with an initiation interval (II) of 1, i.e., one event processed per clock cycle.
\begin{table}[ht!]
\centering
\caption{Algorithmic comparison for FPGA-based event-stream clustering.}
\label{tab:algo_comparison}
\setlength{\tabcolsep}{3pt}
\begin{tabular}{|p{1.4cm}|p{1.4cm}|p{1.4cm}|p{3.5cm}|}
\hline
\textbf{Algorithm} & \textbf{Complexity} & \textbf{Data Paradigm} & \textbf{Suitability for this Work} \\
\hline
K-Means & $O(n \cdot k \cdot i)$ & Batch / static & \textbf{Unsuitable.} Requires known $k$; iterative; multiple passes \cite{Liu2021TiAcc}. \\
\hline
DBSCAN & $O(n^2)$ or $O(n\log n)$ & Batch / static & \textbf{Inefficient.} High memory demand for $\varepsilon$-neighborhood search \cite{Shi2014}. \\
\hline
Grid-Clustering & $O(n)$ & \textbf{Streaming} & \textbf{Ideal.} Linear complexity; direct grid mapping; minimal memory; FPGA-friendly \cite{Schikuta1996}. \\
\hline
\multicolumn{4}{p{215pt}}{Abbreviations: $k$ – number of clusters, $i$ – iterations, $\varepsilon$ – neighborhood radius.} \\
\end{tabular}
\end{table}
\subsection{Grid Clustering Algorithm and Hybrid Data Flow}
\begin{figure}[t]
 \centering
 \includegraphics[width=0.5\textwidth]{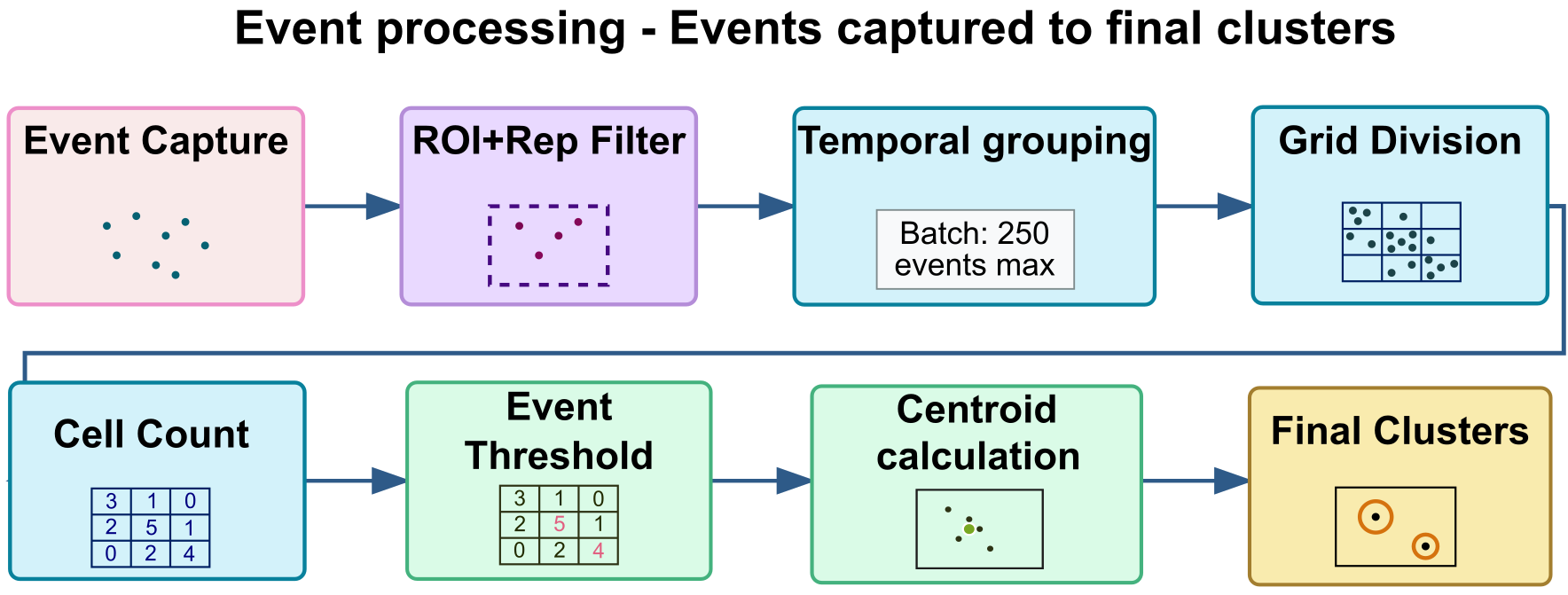}
 \caption{Data/event processing pipeline showing event capture from a single EBC, FPGA grid quantization, and cluster detection stages.}
\label{fig2}
\end{figure}

The core processing algorithm implements a novel, hybrid adaptation of the classical grid clustering method \cite{Schikuta1996}. Fig.~\ref{fig2} depicts the complete data pipeline. The processing is divided into two distinct stages:
\begin{enumerate}
    \item \textbf{Hardware-Accelerated Spatial Quantization (FPGA Server):} The FPGA board, running a dedicated server script, is dedicated exclusively to the massively parallel task of spatial quantization. The overlay receives a stream of $(x, y)$ event coordinates via DMA. The custom IP core efficiently calculates and returns the corresponding grid cell index $(cell\_x, cell\_y)$ for each event based on the fixed $16\times 16$ grid size.
    \item \textbf{Software-Based Cluster Formation (Client):} The embedded client receives the stream of quantized cell indices from the server. It performs the stateful logic of cluster formation: it aggregates events by their $(cell\_x, cell\_y)$ key, counts them to ensure they meet the minimum event threshold, and calculates the final cluster centroid.
\end{enumerate}

This distributed approach leverages the FPGA for the task it excels at—parallel, stateless computation (quantization)—while reducing the load of the sequential, stateful logic (counting and centroid calculation) to the client's CPU.

\subsection{Event-Counting Detection Mechanism}
The system exploits the high dynamic range (120 dB) and temporal responsiveness of the neuromorphic vision sensor to enable detection through asynchronous event accumulation. Unlike conventional frame-based cameras that rely on absolute intensity measurements, the EBC generates sparse data streams where each pixel independently reports relative brightness changes exceeding a predefined contrast threshold. Detection is therefore performed not by analyzing static contrast differences between objects, but by counting and clustering temporally correlated events within spatial regions over sliding time windows.
In this paradigm, contrast sensitivity serves as the triggering mechanism for event generation: when an RSO moves across the sensor's field of view (FoV), it induces localized brightness variations at the pixel level, which in turn produce sequences of events. The detection algorithm aggregates these events within discrete grid cells, applying a minimum event-count threshold to distinguish genuine RSO trajectories from background noise. This event-counting approach provides three key advantages: for space surveillance: (1) immunity to motion blur due to microsecond temporal resolution, (2) robustness against varying illumination conditions since detection depends on relative changes rather than absolute brightness, and (3) computational efficiency by processing only relevant spatiotemporal data instead of full frames.
The grid-based clustering stage reinforces this mechanism by enforcing spatial coherence: events must not only exceed the temporal count threshold but also form continuous patterns consistent with expected orbital motion. This two-stage filter---temporal event accumulation followed by spatial pattern validation---significantly reduces false positives while maintaining high sensitivity to faint, fast-moving targets that would be challenging to detect using conventional contrast-discrimination methods or conventional frame-based cameras.

\subsection{Cluster Quality Metrics}
To quantitatively assess the quality of detected clusters and determine the optimal minimum event threshold, we defined a set of statistical metrics computed on the image region surrounding each cluster. For every detected cluster, we extracted a $48~\times~48$ pixel window centered on its centroid from the corresponding reconstructed frame. Within this window, we computed:

\begin{itemize}
    \item \textbf{Shannon entropy} \cite{ShannonEntropy}: $H = -\sum_{i} p_i \log_2 p_i$, where $p_i$ is the normalized histogram of pixel intensities from the reconstructed frame.
    \item \textbf{Rényi entropy (order 2)} \cite{RenyiEntropy}: $H_2 = -\log_2 \sum_i p_i^2$.
    \item \textbf{Differential entropy} \cite{DiffEntropy}: based on the standard deviation of gradient magnitudes.
    \item \textbf{Local contrast}: the standard deviation of pixel intensities within the window.
    \item \textbf{Edge density}: the ratio of edge pixels (detected by Canny) to total pixels.
    \item \textbf{Event count}: the number of events that contributed to the cluster.
\end{itemize}

These metrics were computed for all clusters across the six validation recordings (each with three lens configurations). The analysis revealed that clusters with fewer than five events rarely correspond to genuine RSOs, establishing the optimal threshold of $min\_events = 5$ used throughout this work.

\section{Implementation}
\label{sec:implementation}

\subsection{Hardware Overlay Design}
\begin{figure}[h!]
 \centering
 \includegraphics[width=0.5\textwidth]{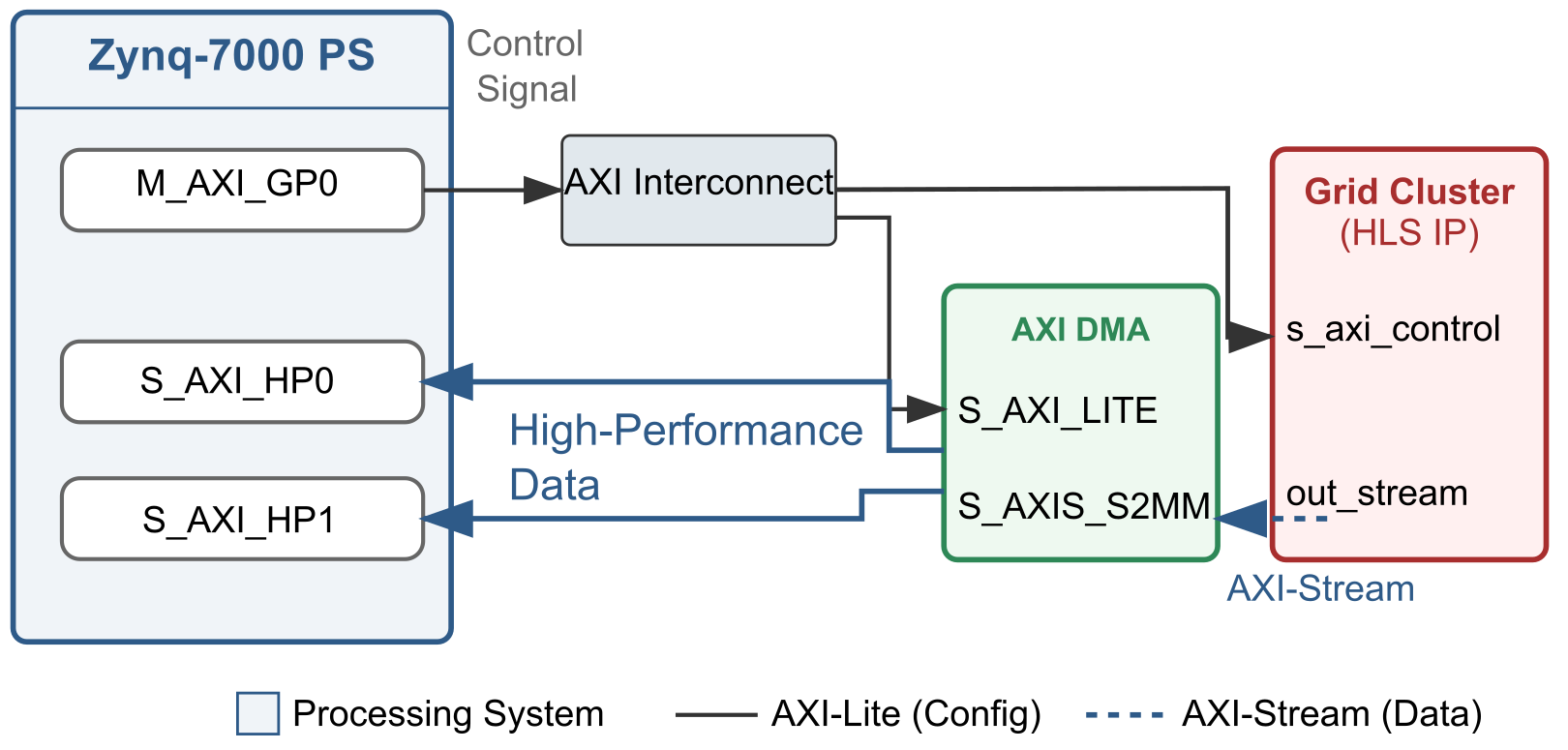}
 \caption{Hardware block diagram showing the general FPGA overlay implementation with DMA controllers and processing system integration.}
\label{figVivado}
\end{figure}

The FPGA implementation targets a PYNQ-Z2 board featuring a Xilinx Zynq-7020 SoC. The custom overlay, shown in Fig.~\ref{figVivado}, integrates the grid clustering IP core, AXI DMA controllers, and AXI-Lite interfaces for configuration. The processing system (PS) runs a lightweight Python server that manages DMA transfers and communicates with the client over TCP/IP.

Resource utilization on the Zynq-7020 is summarized in Table~\ref{tab:resources}. The design leaves ample capacity for future extensions such as on-chip tracking or neural network accelerators.

\begin{table}[h!]
\centering
\caption{FPGA resource utilization for grid clustering overlay.}
\label{tab:resources}
\begin{tabular}{lccc}
\hline
\textbf{Resource} & \textbf{Used} & \textbf{Available} & \textbf{Utilization (\%)} \\
\hline
LUTs & 34,560 & 53,200 & 65\% \\
Flip-Flops & 42,240 & 106,400 & 40\% \\
DSP Blocks & 121 & 220 & 55\% \\
BRAM (36 Kb) & 84 & 140 & 60\% \\
\hline
\end{tabular}
\end{table}

\subsection{Microarchitecture of the Grid Clustering IP Core}

\begin{figure*}[htbp]
 \centering
 \includegraphics[width=1.0\textwidth]{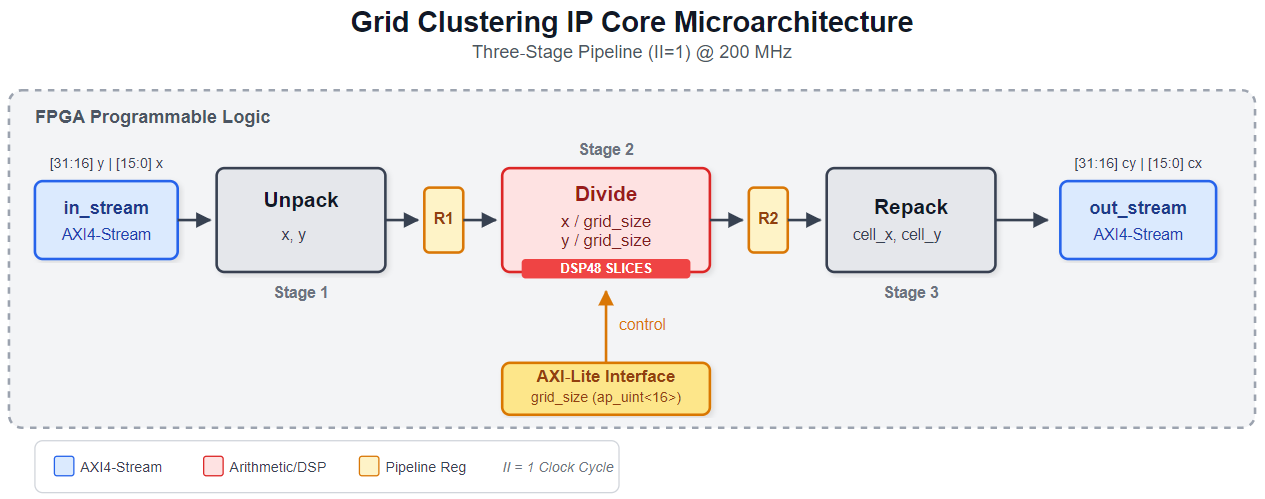}
 \caption{Microarchitecture of the grid clustering IP core implemented in HLS. The core extracts $(x,y)$ coordinates from AXI-Stream packets, computes cell indices via division, and repacks the result with an initiation interval of 1.}
\label{figIPMicro}
\end{figure*}

The grid clustering IP core is implemented in C++ using Xilinx Vitis HLS, enabling rapid design space exploration while achieving high quality of results. The core's microarchitecture, depicted in Fig.~\ref{figIPMicro}, follows a fully pipelined streaming paradigm. It accepts a continuous stream of 32-bit AXI4-Stream packets, where the lower 16 bits encode the $x$ coordinate and the upper 16 bits encode the $y$ coordinate of an event. The processing stages are:

\begin{enumerate}
    \item \textbf{Input Interface:} The core monitors the incoming AXI4-Stream FIFO. When a valid packet is present, it reads the 32-bit data word.
    \item \textbf{Coordinate Extraction:} The $x$ and $y$ fields are extracted using bit-slicing operations: $x = data(15,0)$ and $y = data(31,16)$.
    \item \textbf{Grid Quantization:} The grid cell indices are computed as $cell\_x = x / grid\_size$ and $cell\_y = y / grid\_size$, where $grid\_size$ is a configurable parameter provided via an AXI-Lite register. The division operations are synthesized using the FPGA's DSP48 slices for high throughput.
    \item \textbf{Output Packing:} The resulting $(cell\_x, cell\_y)$ pair is packed back into a 32-bit word, with $cell\_x$ occupying bits 15–0 and $cell\_y$ occupying bits 31–16.
    \item \textbf{Output Interface:} The packet is driven onto the output AXI4-Stream interface, preserving the handshake signals.
\end{enumerate}

The use of the \texttt{HLS PIPELINE} directive enables an initiation interval (II) of 1, meaning the core can accept a new input event every clock cycle. At the nominal operating frequency of 200 MHz, this yields a peak throughput of 200 million events per second—well beyond the requirements of current neuromorphic sensors. The division latency is fully absorbed by the pipeline stages, ensuring that the overall system latency is dominated by data transfer and software processing rather than the IP core itself.

\subsection{Client-Server Communication Protocol}
The client and server communicate over a TCP socket using a custom binary protocol. The client serializes batched event coordinates using Python Pickle (protocol version 2) and prefixes each message with a 4-byte length header. The server deserializes the events, transfers them to the FPGA via DMA, and returns the quantized cell indices. The total round-trip time, including serialization, network transfer, DMA operations, and FPGA processing, is consistently below 25 ms for batches of 250 events.
To ensure robustness, the client implements a send queue and non-blocking receives, allowing the event capture thread to continue operation even if the FPGA server experiences transient delays. The server, in turn, can handle multiple client connections using threaded socket handling.

\subsection{Performance Optimization}
Performance optimization focuses on minimizing end-to-end latency while maintaining power efficiency. The complete processing pipeline achieves total system latencies below 62 ms, measured from event arrival at the client to cluster centroid output. A breakdown of this latency is provided in Table~\ref{tab:latency}. The FPGA acceleration component contributes less than 25 ms, while network transfer and software clustering account for the remainder.

\begin{table}[h!]
\centering
\caption{Latency breakdown per processing stage (batch size 250 events).}
\label{tab:latency}
\begin{tabular}{lc}
\hline
\textbf{Stage} & \textbf{Latency (ms)} \\
\hline
Event accumulation (20 ms window) & 20.0 \\
Serialization and TCP send & 2.1 \\
FPGA DMA transfer + quantization & 0.8 \\
TCP receive and deserialization & 1.5 \\
Software clustering (aggregation) & 12.3 \\
Visualization and tracking & 25.0 \\
\hline
\textbf{Total} & \textbf{61.7} \\
\hline
\end{tabular}
\end{table}

Power optimization employs clock gating and dynamic frequency scaling. The complete system demonstrates total power consumption of 8.5 W during continuous operation, with the FPGA component consuming 2.3 W and the camera consuming 1 W.

\section{Experimental Validation and Results}
\label{sec:results}

\subsection{Experimental Setup}
System validation employed a comprehensive multi-stage approach. Laboratory validation utilized the EVAS dataset \cite{Valdivia2023} containing labeled RSO trajectories. We selected six representative recordings, each captured with three distinct lens types (standard, telephoto, and wide-angle). Field tests were conducted at the SeeTRUE observatory\cite{SEETRUE} in Valparaíso, Chile under various night sky conditions. Simultaneous telescope observations served as ground truth, with angular velocities up to 0.6 rad/s. Power measurements were taken using calibrated USB power meters.
Detection accuracy was evaluated through systematic sampling of 1,000 detection events across all six recordings. Each cluster was manually verified against corresponding telescope imagery; a true positive was recorded if the cluster centroid coincided with a known RSO trajectory. This process yielded the 97\% accuracy reported in Table~\ref{tab1}.

\begin{table}[h!]
\centering
\caption{System performance specifications and experimental results}
\begin{tabular}{lcc}
\hline
\textbf{Parameter} & \textbf{Value} & \textbf{Unit} \\
\hline
Total system latency & $< 62$ & ms \\
Total power consumption & 8.5 & W \\
FPGA power consumption & 2.3 & W \\
EBC power consumption & 1.0 & W \\
Detection accuracy & 97 & \% \\
Dynamic range & 120 & dB \\
Clustering grid size & $16\times 16$ & cells \\
Minimum events per cluster & 5 & events \\
Event batch size & 250 & events \\
\hline
\end{tabular}
\label{tab1}
\end{table}

\subsection{Statistical Analysis of Detected Clusters}
\begin{figure}[t!]
 \centering
 \includegraphics[width=0.5\textwidth]{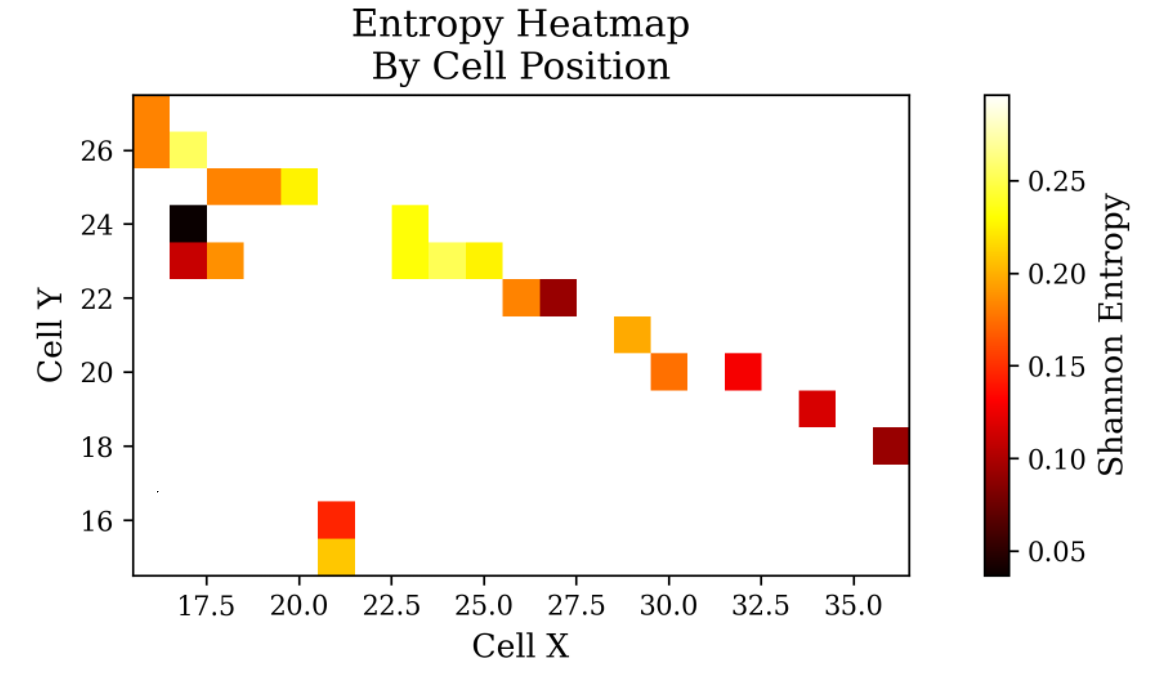}
 \caption{Shannon entropy per grid cell.}
\label{fig5ShannonCell}
\end{figure}

\begin{figure}[t!]
 \centering
 \includegraphics[width=0.5\textwidth]{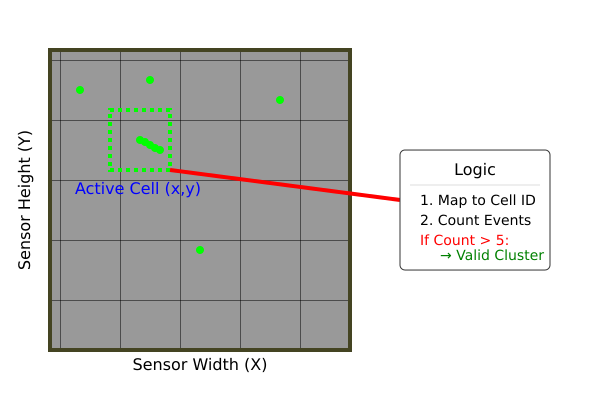}
 \caption{Distribution of events per cluster. The dashed box indicates a valid cluster of at least 5 events.}
\label{figValidCluster}
\end{figure}

\begin{figure}[htbp!]
 \centering
 \includegraphics[width=0.5\textwidth]{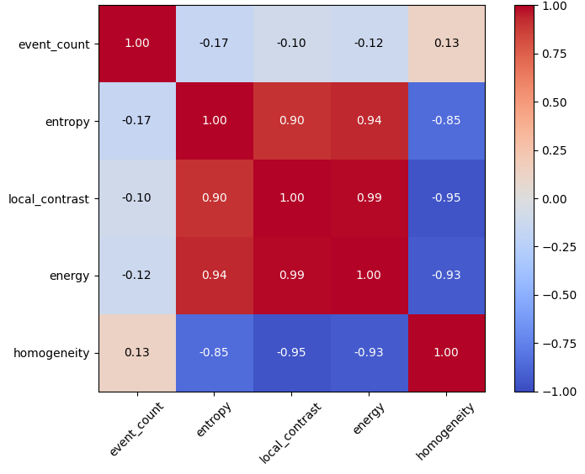}
 \caption{Correlation matrix of cluster quality metrics. Strong positive correlations exist between entropy, contrast, and event count.}
\label{figCorrelation}
\end{figure}

Fig.~\ref{fig5ShannonCell} shows the distribution of Shannon entropy per grid cell for genuine RSOs versus noise clusters (stars). Genuine RSOs exhibit significantly higher contrast and  higher entropy, indicating richer spatial structure. Fig.~\ref{figValidCluster} plots the event count per cluster; the majority of true clusters contain between 5 and 20 events, consistent with our selected threshold. The correlation matrix (Fig.~\ref{figCorrelation}) reveals that entropy metrics are strongly correlated with event count and contrast, confirming their utility as quality indicators.

\begin{figure}[h!]
 \centering
 \includegraphics[width=0.5\textwidth]{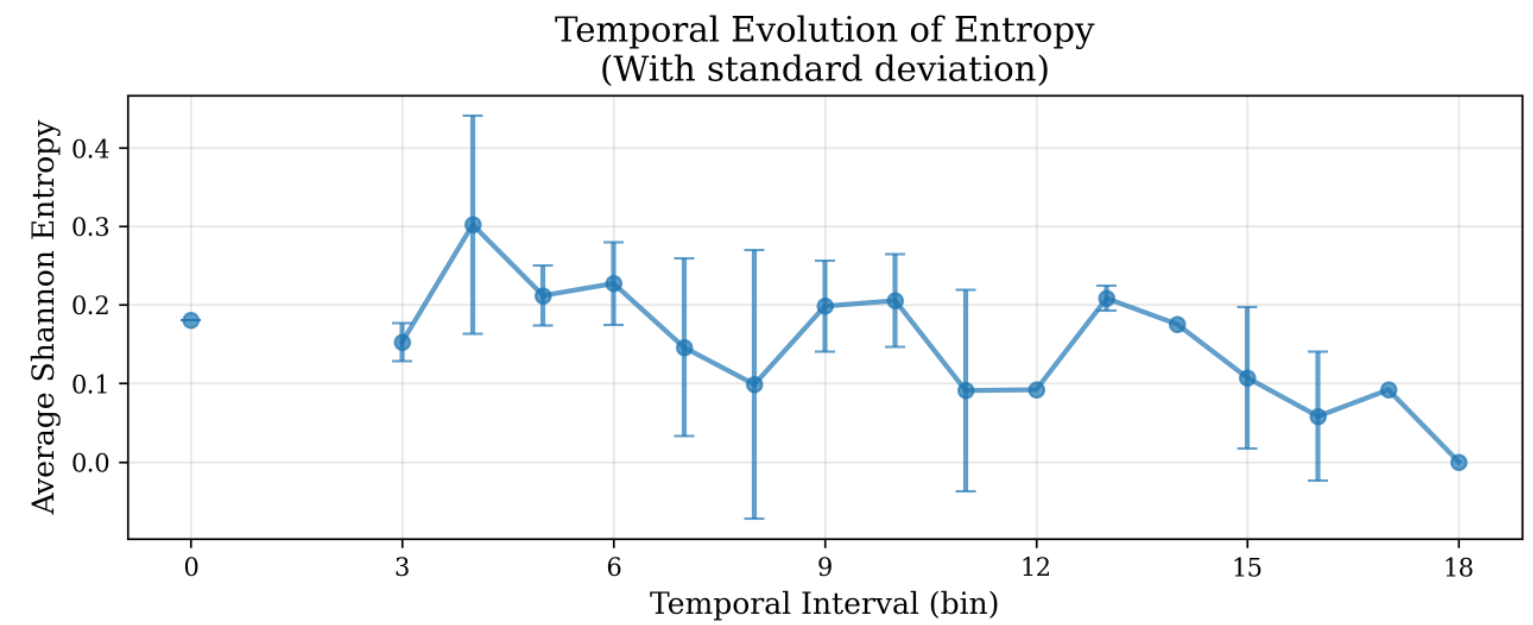}
 \caption{Temporal evolution of Shannon entropy for a tracked RSO over 50 consecutive frames.}
\label{fig4ShannonTime}
\end{figure}

Temporal analysis (Fig.~\ref{fig4ShannonTime}) tracks the average entropy of a RSO cluster across 50 consecutive frames. While noise clusters (recordings of fixed stars) show erratic fluctuations, the tracked object maintains a stable entropy profile, confirming the consistency of our tracking logic.

\begin{figure}[h!]
 \centering
 \includegraphics[width=0.5\textwidth]{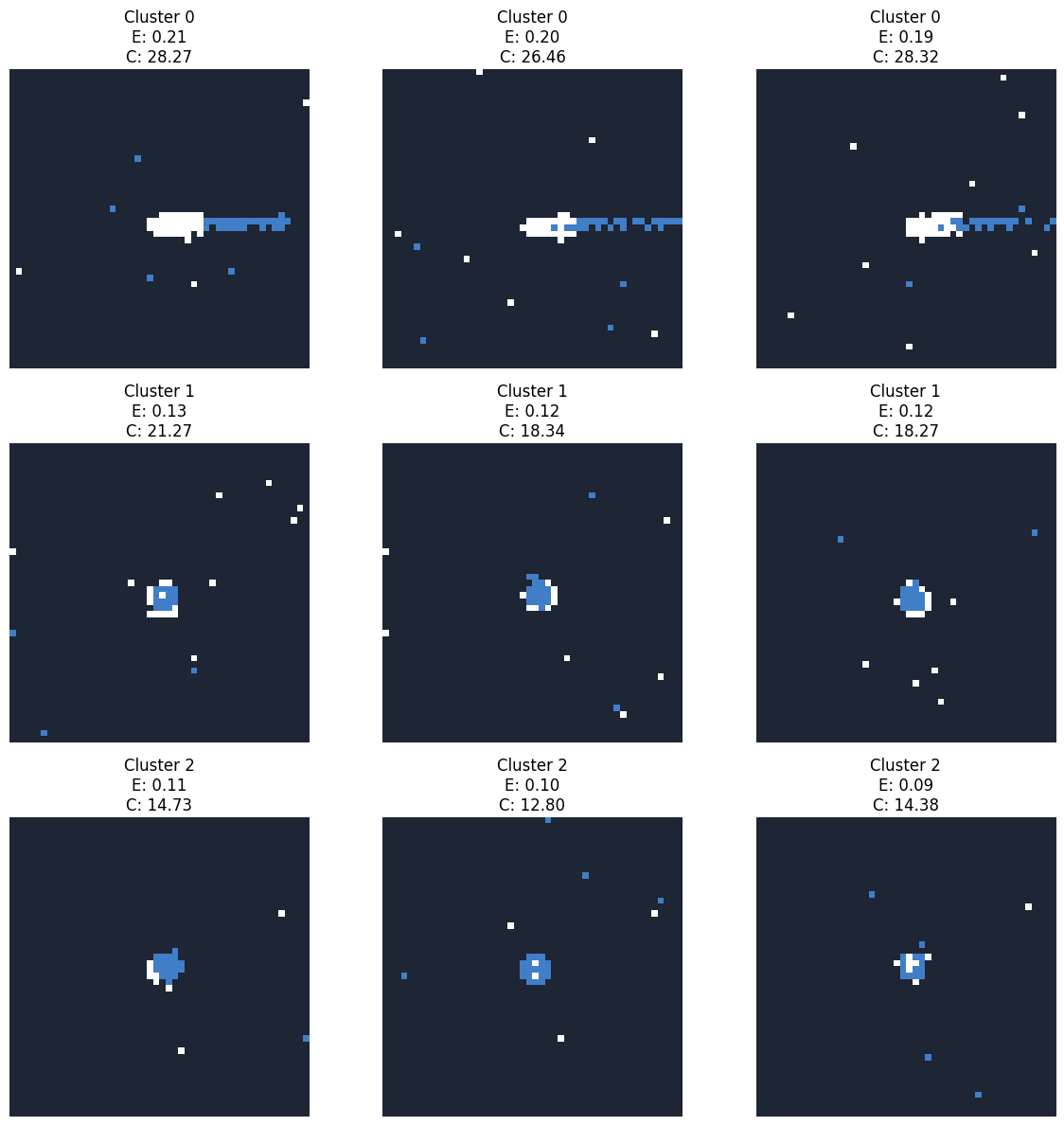}
 \caption{Example of segmented objects captured from night sky surveillance data, illustrating temporal instances of tracked objects and entropy-based visual differences between moving satellites and stars. Each example presents Contrast (C) and Entropy (E).}
\label{figDetections}
\end{figure}

In Fig.~\ref{figDetections}, objects captured by the complete system are shown from an example night-sky surveillance recording. The objects are segmented into 48 × 48 pixel patches to observe their local surroundings. The same object, captured at different temporal instances, can be observed along the same row. Higher-entropy objects tend to resemble moving satellites, whereas lower-entropy objects correspond to stars exhibiting apparent motion due to the Earth’s rotation and the passage of time.

\subsection{Performance Benchmarks and Power Efficiency}
\begin{figure*}[h!]
 \centering
 \includegraphics[width=\textwidth]{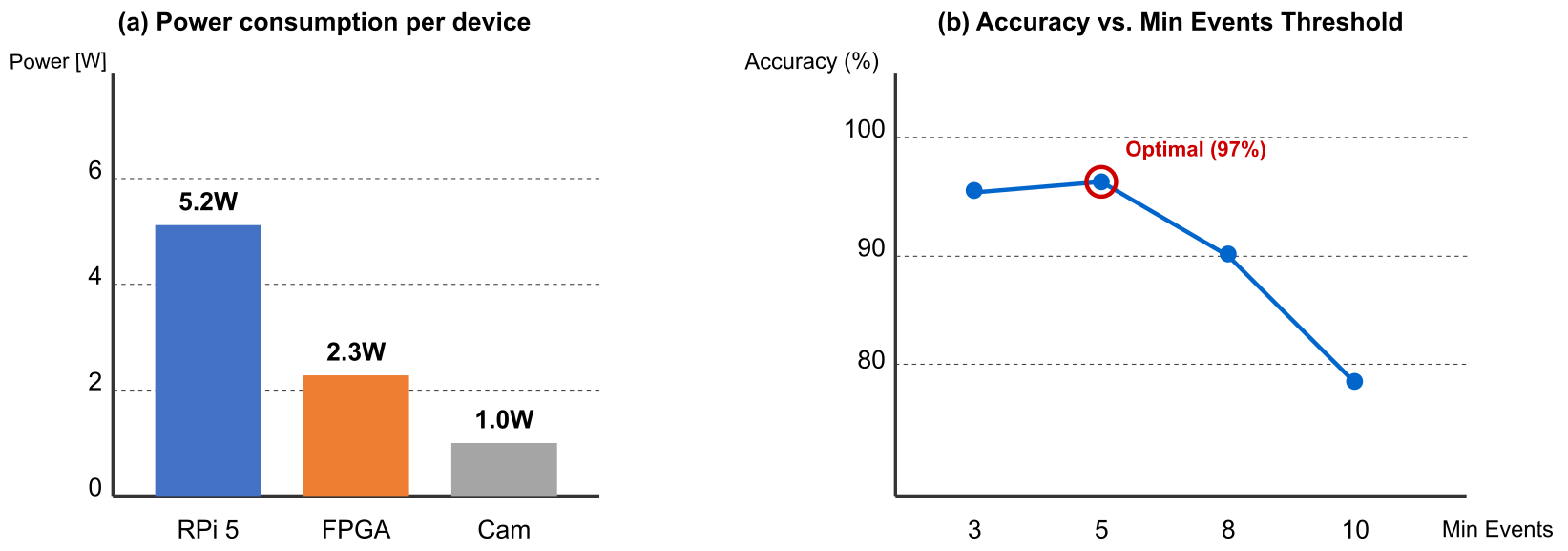}
 \caption{(a) Power consumption breakdown of the complete system. (b) Detection accuracy versus minimum events per cluster threshold. The optimal threshold is 5 events, yielding 97\% accuracy.}
 \label{fig10}
\end{figure*}

Fig.~\ref{fig10} presents the system's power consumption breakdown and detection accuracy as a function of the minimum events threshold. The optimal operating point is at $min\_events = 5$, achieving 97\% accuracy. The total power consumption of 8.5 W is dominated by the embedded processor (client) and peripheral components; the FPGA contributes only 2.3 W, demonstrating the efficiency of hardware acceleration.

\subsection{Field Deployment and Scalability}
\label{sec:results:field}
\begin{figure}[h!]
 \centering
 \includegraphics[width=0.5\textwidth]{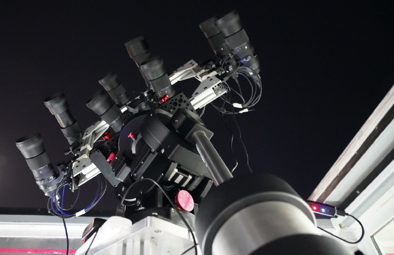}
 \caption{Multiple-EBC configuration (ARACHNID \cite{ARACHNID}) with narrow-FoV lenses, providing continuous coverage from 45° to 90° elevation.}
\label{fig5}
\end{figure}

\begin{figure}[h!]
 \centering
 \includegraphics[width=0.5\textwidth]{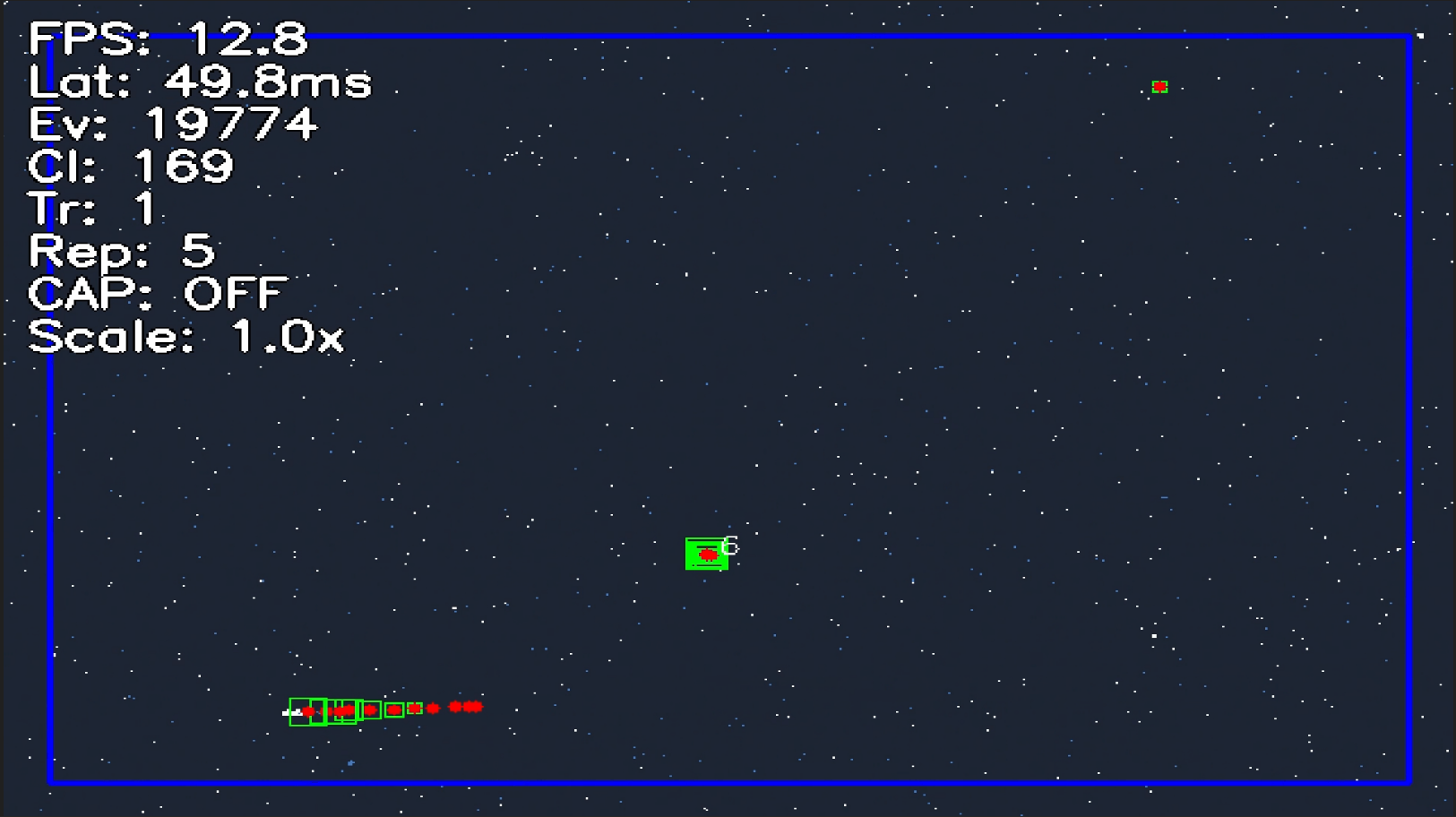}
 \caption{Example of detections on the implemented single-EBC FPGA setup during field tests.}
\label{fig6}
\end{figure}

Fig.~\ref{fig5} illustrates the ARACHNID multi-EBC configuration designed for continuous all-sky coverage. Each EBC is paired with a dedicated FPGA board, forming an independent processing node. Fig.~\ref{fig6} shows an example detection from the single-EBC field deployment. The modular architecture ensures that adding a new EBC-FPGA unit increments the total power budget by approximately 3.3 W while preserving deterministic latency below 62 ms per stream.

\subsection{Scalability Analysis for Multi-Camera Deployments}
The modular design of the system enables straightforward scaling by adding identical EBC–FPGA nodes. Table~\ref{tab:scalability} quantifies the aggregate performance metrics for configurations ranging from one to eight EBCs. The total power consumption scales linearly with the number of nodes, as each FPGA contributes 2.3 W and each EBC contributes 1.0 W. The peak event throughput scales nearly linearly, limited only by the Gigabit Ethernet switch capacity. Crucially, the per-EBC processing latency remains invariant at approximately 62 ms, demonstrating that the architecture avoids resource contention and preserves deterministic timing even under multi-stream operation.

\begin{table}[h!]
\centering
\caption{Scalability analysis for multi-EBC deployments.}
\label{tab:scalability}
\begin{tabular}{lcccc}
\hline
\textbf{Metric} & \textbf{1 EBC} & \textbf{2 EBCs} & \textbf{4 EBCs} & \textbf{8 EBCs} \\
\hline
Total power (W)           & 8.5  & 11.8 & 18.4 & 31.6 \\
Peak throughput (kEv/s)   & 250  & 495  & 990  & 1980 \\
Network bandwidth (Mbps)  & 12.5 & 25.0 & 50.0 & 100.0 \\
\hline
\end{tabular}
\end{table}

The predictable power and latency scaling make the system suitable for deployment in remote observatories or on small satellite constellations, where power budgets are strictly limited and real-time responsiveness is mandatory. The use of commercial off-the-shelf components further reduces deployment costs and simplifies maintenance.

\section{Discussion}
\label{sec:discussion}
The experimental results validate the effectiveness of the proposed hybrid architecture. The 97\% detection accuracy, achieved with a simple grid clustering algorithm and a fixed threshold of 5 events, underscores the synergy between neuromorphic sensing and hardware acceleration. The entropy-based analysis not only justifies the chosen threshold but also provides a quantitative framework for tuning the system to different environmental conditions.
The system's 62 ms latency is dominated by software stages (event accumulation, clustering, and visualization). Future optimizations could offload the aggregation and centroid calculation to the FPGA fabric, potentially reducing total latency to below 30 ms. Nevertheless, the current performance already meets the requirements for tracking LEO satellites and meteors.
The use of commercial off-the-shelf components and an open-source implementation lowers the barrier to adoption for space surveillance and other real-time event-based applications. The modular architecture facilitates scaling to multi-EBC networks, as demonstrated by the ARACHNID concept \cite{ARACHNID}, enabling cost-effective all-sky monitoring.

\section{Conclusion}
\label{sec:conclusion}
This work has presented a comprehensive implementation of an FPGA-accelerated neuromorphic vision system for real-time detection of resident space objects (RSOs), optimizing the grid clustering algorithm \cite{Schikuta1996} for space surveillance. The system integrates a single event-based camera (EBC) with a custom hardware overlay that executes spatial quantization in programmable logic, achieving total latency below 62 ms and power consumption of 8.5 W. Experimental validation demonstrated 97\% detection accuracy for targets with angular velocities up to 0.6 rad/s, using an optimal threshold of 5 events per cluster determined through information-theoretic analysis.
The architecture provides a scalable foundation for multi-EBC deployments, where each additional EBC–FPGA unit adds approximately 3.3 W while maintaining deterministic performance. A significant secondary benefit is the automated generation of annotated datasets for training neural networks, as each detection is inherently timestamped and spatially localized.
Future work will focus on deploying the ARACHNID multi-EBC array for continuous all-sky coverage, integrating more sophisticated tracking algorithms, and exploring on-FPGA spiking neural network implementations for object classification.

\section*{Acknowledgment}
The authors declare no conflict of interest. The experimental data used for validation was sourced from the publicly available EVAS dataset \cite{Valdivia2023}.

The authors disclose the use of artificial intelligence (AI) tools during the preparation of this work. The DeepSeek-Coder 6.7B model (via Ollama) \cite{OllamaDeepSeek} was used to assist in generating and optimizing the Python source code for both the server infrastructure and the client display module. Text editing and grammar enhancement for selected subsections within the manuscript were supported by the \texttt{Gemini-Flash} model.

\bibliographystyle{IEEEtran}
\bibliography{references}

\begin{IEEEbiography}[{\includegraphics[width=1in,height=1.25in,clip,keepaspectratio]{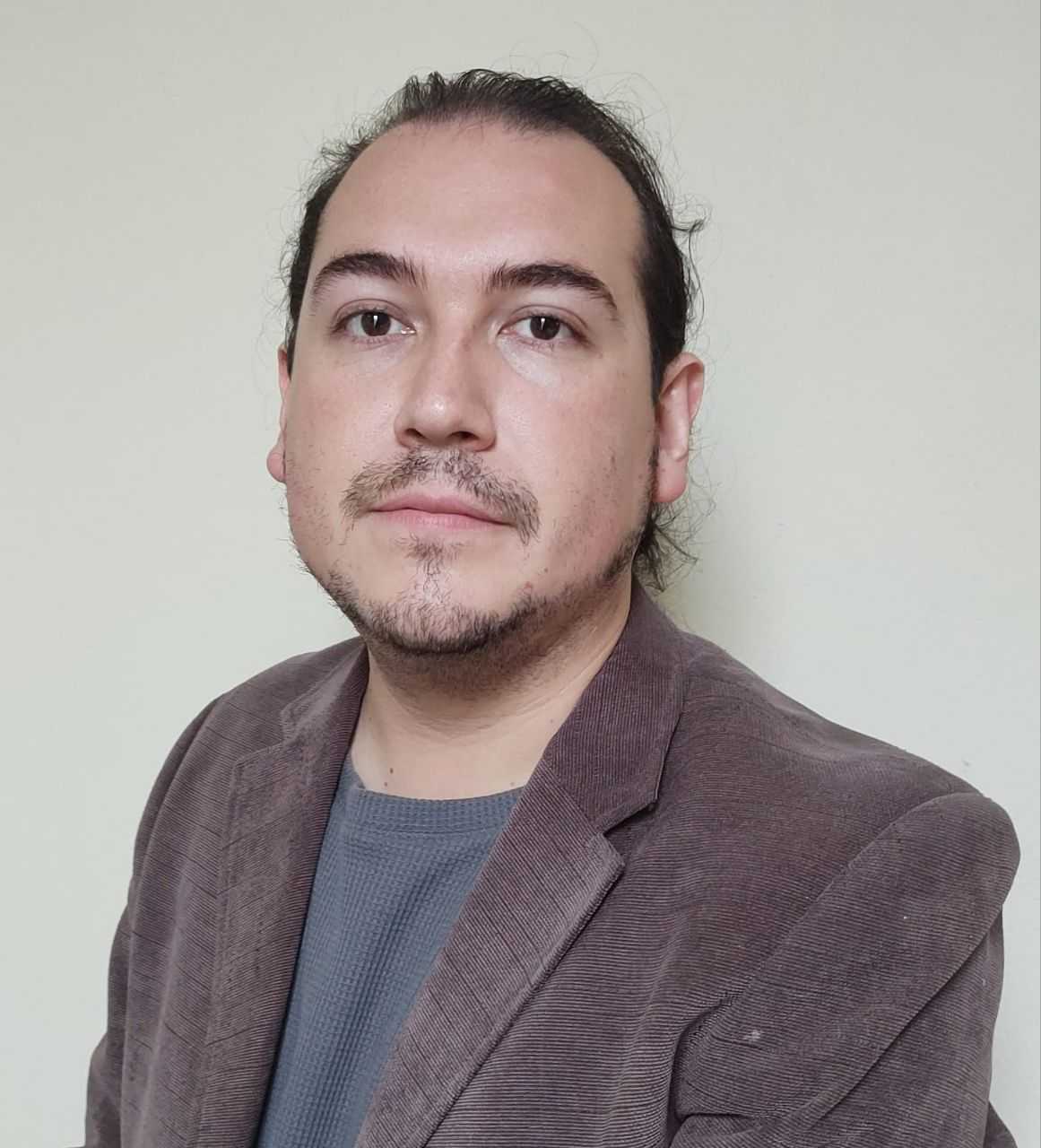}}]{Diego Hern\'{a}ndez}
Diego Hernández received the B.Sc. degree in electronic engineering and the M.Sc. degree in electrical engineering from the Pontificia Universidad Católica de Valparaíso (PUCV), Valparaíso, Chile, in 2022 and 2025, respectively. He was a Researcher at the Optoelectronics Laboratory (Optolab), PUCV, where he contributed to neuromorphic vision systems and IoT prototypes. Simultaneously, he worked as a Project Engineer at LabSens, PUCV, developing high-speed machine vision solutions for industrial quality control. In 2024, he joined SeeTRUE, Curauma, Chile, as a Project and Research Engineer, where he currently leads the development of FPGA-accelerated real-time vision systems for orbital object detection and space situational awareness. His research interests include event-based vision, FPGA acceleration, edge artificial intelligence, and low-power embedded systems.
\end{IEEEbiography}

\begin{IEEEbiography}[{\includegraphics[width=1in,height=1.25in,clip,keepaspectratio]{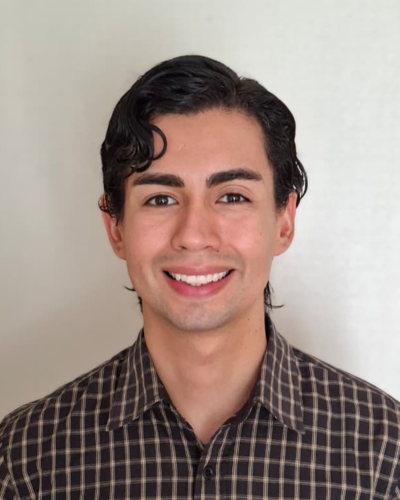}}]{Sebasti\'{a}n Valdivia} received the B.Sc. degree in electronic engineering and the M.S. degree in electrical engineering from the Pontificia Universidad Cat\'{o}lica de Valpara\'{i}so (PUCV), Valparaíso, Chile, in 2022 and 2025, respectively. From 2022 to 2025, he was a Researcher with the Optoelectronics Laboratory, PUCV, where he worked on neuromorphic vision systems for space situational awareness, including the ARACHNID system. He is currently a Researcher with LabSens, PUCV, where he develops event-based and embedded systems using spiking neural networks. His research interests include neuromorphic vision, space situational awareness, event-driven sensing, and low-power intelligent systems.
\end{IEEEbiography}

\begin{IEEEbiography}[{\includegraphics[width=1in,height=1.25in,clip,keepaspectratio]{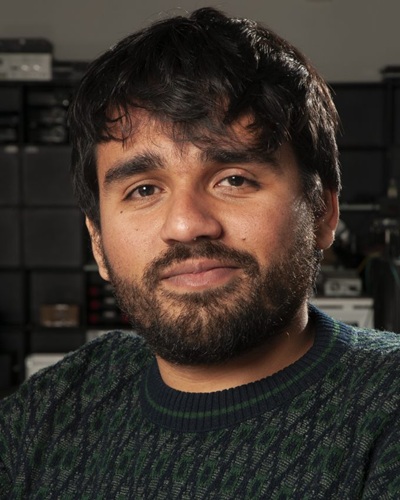}}]{Vicente Westerhout} received the B.S./M.S. degrees in Electrical Engineering from Pontificia Universidad Cat\'{o}lica de Valpara\'{i}so (PUCV) in 2021 and 2023, and has been pursuing the Ph.D. degree in Electrical Engineering there since 2023. His research interests include space situational awareness and atmospheric characterization using neuromorphic vision systems.
\end{IEEEbiography}

\begin{IEEEbiography}[{\includegraphics[width=1in,height=1.25in,clip,keepaspectratio]{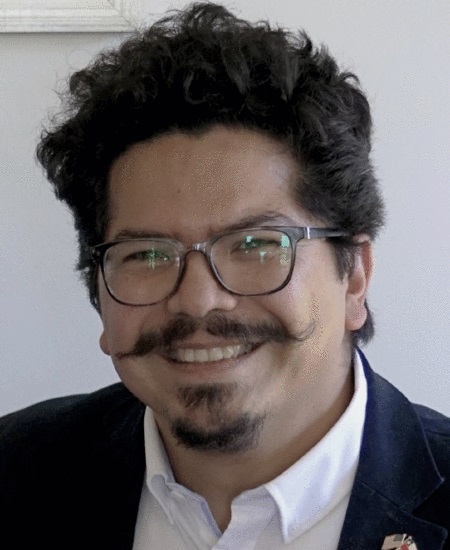}}]{Esteban Vera} (SM) received the engineering diploma in electronics engineering and the M.S. and Ph.D. degrees in electrical engineering from Universidad de Concepci\'{o}n, Concepci\'{o}n, Chile, in 1999, 2003, and 2010, respectively.
From 2001 to 2007, he worked for the Paranal and Gemini Observatories. In 2010, he was a Postdoctoral Researcher with the University of Arizona, Tucson, AZ, USA, and then a Research Scientist with Duke University, Durham, NC, USA, from 2013. In 2016, he joined the School of Electrical Engineering, Pontificia Universidad Cat\'{o}lica de Valpara\'{i}so, Chile, where he is an Associate Professor and leads the Optoelectronics Lab (Optolab). His research interests include computational imaging, compressed sensing, machine learning, optical computing, and adaptive optics.
Dr. Vera is an Associate Editor for IEEE Transactions on Computational Imaging, IEEE Open Journal of Signal Processing, and Optics Express. He is a Member of the SPIE and a Senior Member of OPTICA.
\end{IEEEbiography}

\begin{IEEEbiography}[{\includegraphics[width=1in,height=1.25in,clip,keepaspectratio]{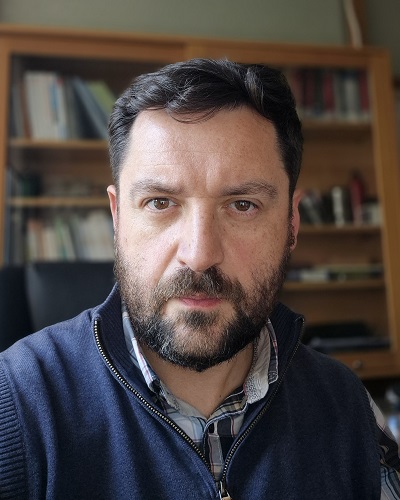}}]{Daniel Yunge} (M'08) was born in Valdivia, Chile, in 1984. He received the Electronics Engineering degree from the Pontificia Universidad Católica de Valparaíso (PUCV), Chile, in 2010, and the Ph.D. degree in Elektrotechnik from the Technical University of Munich (TUM), Germany, in 2019. Since 2022, he has been an Assistant Professor and co-head of LabSens at PUCV, where he focuses on real-time embedded and sensing systems, as well as energy-constrained AI systems, including neuromorphic computing and TinyML, for applications in aerospace, agriculture, and healthcare.
\end{IEEEbiography}

\end{document}